\documentclass[%
 reprint,
 amsmath,amssymb,
 aps,
 prl,
]{revtex4-2}

\usepackage{aas_macros}
\usepackage{graphicx}% Include figure files
\usepackage{dcolumn}% Align table columns on decimal point
\usepackage{bm}% bold math
\usepackage{hyperref}% add hypertext capabilities
\usepackage{color}
\usepackage{natbib}
\usepackage{CJK}
\usepackage[english]{babel}

\newcommand{\jcapp}{{J.~Cosmol.~Astropart.~Phys.}}

\begin{document}
\begin{CJK*}{UTF8}{gbsn} % Use default fonts from CJK (see below)
\preprint{APS/123-QED}

\title{Neutrino Fast Flavor Conversions in Neutron-star Post-Merger Accretion Disks}% Force line breaks with \\
%\thanks{A footnote to the article title}%

\author{Xinyu Li (李昕宇)$^{1,2}$}
\email{xli@cita.utoronto.ca}

\author{Daniel M.~Siegel$^{1,3}$}
 
\affiliation{$^1$Perimeter Institute for Theoretical Physics, Waterloo, Ontario, Canada, N2L 2Y5\\
$^2$Canadian Institute for Theoretical Astrophysics, Toronto, Ontario, Canada, M5R 2M8\\
$^3$Department of Physics, University of Guelph, Guelph, Ontario, Canada, N1G 2W1} %\\%Lines break 

%\collaboration{MUSO Collaboration}
%\noaffiliation

%\date{\today}% It is always \today, today,
             %  but any date may be explicitly specified

\begin{abstract} 
A compact accretion disk may be formed in the merger of two neutron stars or of a neutron star and a stellar-mass black hole. 
Outflows from such accretion disks have been identified as a major site of rapid neutron-capture (r-process) nucleosynthesis and as the source of `red' kilonova emission following the first observed neutron-star merger GW170817. 
We present long-term general-relativistic radiation magnetohydrodynamic simulations of a typical post-merger accretion disk at initial accretion rates of $\dot{M}\sim 1\,M_\odot\,\text{s}^{-1}$ over 400\,ms post-merger. 
We include neutrino radiation transport that accounts for effects of neutrino fast flavor conversions dynamically.
We find ubiquitous flavor oscillations that result in a significantly more neutron-rich outflow, providing lanthanide and 3rd-peak r-process abundances similar to solar abundances. 
This provides strong evidence that post-merger accretion disks are a major production site of heavy r-process elements. A similar flavor effect may allow for increased lanthanide production in collapsars.
%The formalism presented here may also be used in simulations of core-collapse supernovae to explore whether fast conversions strengthen or weaken the explosion.
\end{abstract}

%\keywords{Suggested keywords}%Use showkeys class option if keyword
\maketitle
\end{CJK*}
%%%%%%%%% Introduction %%%%%%%%%

\textit{Introduction.---}Since their first detection \cite{cowan_detection_1956}, neutrinos have yielded surprises in fundamental physics, astrophysics, and cosmology. 
The discovery of vacuum oscillations \cite{fukuda_evidence_1998,ahmad_measurement_2001} provided a solution to the long-standing solar neutrino problem \cite{bahcall_are_1997,ahmad_measurement_2001,fukuda_solar_2001}. Oscillations are altered by the presence of electrons in matter (the Mikheyev-Smirnov-Wolfenstein effect) \cite{wolfenstein_neutrino_1978,mikheyev_resonance_1985,mikheev_neutrino_1986}, relevant for atmospheric and solar neutrinos \cite{fukuda_evidence_1998,fukuda_solar_2001}.
In astrophysical systems with dense neutrino gases or the early universe, neutrinos themselves may act as a refractive medium instead of electrons, leading to self-induced flavor conversions \cite{pantaleone_neutrino_1992,sigl_general_1993,samuel_neutrino_1993}. 
Given seed perturbations \cite{sawyer_neutrino_2016}, such conversions can develop run-away modes in flavor space on a timescale $\Phi_0 = \sqrt{2}G_F \hbar^{-1} n_{\nu} = 1.92\times 10^9\,\mathrm{s}^{-1} (n_{\nu}/10^{31}\,\mathrm{cm}^{-3})$ much faster than vacuum oscillations \cite{sawyer_speed-up_2005,chakraborty_collective_2016,Chakraborty_selfinduced_2016,dasgupta_fast_2017}.

In dense astrophysical environments, neutrinos change the composition and dynamics of matter via the charged-current interactions $\bar{\nu}_e + p \rightleftharpoons e^+ + n $ and $\nu_e + n \rightleftharpoons e^- + p$. 
Since typically $\nu_e$ and $\bar{\nu}_e$ are more copiously produced than their $\mu$ and $\tau$ counterparts, fast flavor conversions in the decoupling region above the neutrinosphere can impact the dynamics of and nucleosynthesis in the outflowing plasma.
Fast conversions have mostly been studied in the context of core-collapse supernovae with focus on their impact on the explosion mechanism \cite{pastor_flavor_2002,sawyer_speed-up_2005,duan_coherent_2006,izaguirre_fast_2017,duan_collective_2010,chakraborty_collective_2016}, but their relevance to neutron-star mergers has been appreciated as well \cite{wu_fast_2017,wu_imprints_2017}.

The merger of two neutron stars in a binary system typically leads to the formation of a compact, neutrino-cooled accretion disk around a remnant black hole \cite{ruffert_coalescing_1997,shibata_merger_2006}. 
Dense neutron-rich outflows from such disks may dominate other types of ejecta from the collision itself and represent a major source of rapid neutron-capture (r-process) elements in the universe \cite{siegel_three-dimensional_2017,kasen_origin_2017,metzger_kilonovae_2019,siegel_gw170817_2019}. 
In particular, neutron-rich outflows (mean proton fraction $Y_e < 0.25$) from such an accretion disk were identified as the source of heavy r-process elements and the associated `red' (lanthanide-bearing) kilonova from the first detected neutron-star merger GW170817 \cite{abbott_gw170817:_2017,abbott_multi-messenger_2017,metzger_kilonovae_2019,siegel_gw170817_2019,radice_dynamics_2020}.

Previous simulations of core-collapse supernovae and neutron-star mergers considered fast conversions either in post-processing \cite{wu_imprints_2017,george_fast_2020,delfan_azari_linear_2019,nagakura_fast-pairwise_2019}, evaluating whether a %given state of the 
system is susceptible to such instabilities, or dynamically, but restricting all calculations to one spatial dimension \cite{stapleford_coupling_2020}. 
Here, we present 3D general-relativistic magnetohydrodynamic simulations with neutrino transport, which include the feedback of fast flavor conversions onto the three-dimensional non-linear dynamics and nucleosynthesis of accretion disks in neutron-star mergers.

%%%%%%%%% Computational setup %%%%%%%%%

\textit{Computational setup.---}We simulate the long-term evolution of a typical post-merger accretion disk with the GRMHD code described in \cite{siegel_three-dimensional_2018}. 
It represents an evolved version of \texttt{GRHydro} \cite{mosta_grhydro:_2014} and makes use of the \texttt{Einstein Toolkit} \footnote{\href{http://einsteintoolkit.org}{http://einsteintoolkit.org}} \citep{maria_babiuc-hamilton_einstein_2019,loffler_einstein_2012,schnetter_evolutions_2004,goodale_cactus_2003,thornburg_fast_2004}. 
We use the SFHo equation of state \cite{2013ApJ...774...17S} as tabulated in Ref.~\cite{oconnor_new_2010} and employ neutrino-matter interactions as implemented by the open-source library \texttt{NuLib} \cite{2015ApJS..219...24O}. 
The neutrino radiation transport follows the general-relativistic moment formalism by Refs.~\cite{thorne_relativistic_1981,shibata_truncated_2011} and is implemented via a multi-group two-moment (M1) scheme, which we adapt from the open-source code of Ref.~\cite{2016ApJ...831...98R}. 
In contrast to Ref.~\cite{2016ApJ...831...98R}, which discarded velocity dependent terms, we retain these, but apply a small-velocity approximation for high-energy neutrinos when the absorption opacity for these neutrinos becomes excessively large; this only affects our highest-energy bin. 
Appropriate closure relations are obtained by the interpolation approach of Ref.~\cite{shibata_truncated_2011} with the standard Minerbo closure \cite{minerbo_maximum_1978}. 
We track the evolution of four neutrino species ($\nu_e$, $\nu_X$, $\bar{\nu}_e$, $\bar{\nu}_X$, where $X$ combines $\mu$ and $\tau$ neutrinos), each with six energy bins equally spaced between $0-60$~MeV. 

Our initial data is given by an equilibrium torus \cite{2011IJMPD..20.1251S} with $M_{t0}=0.07M_\odot$ around a spinning black hole of mass $M_{\rm BH}=3M_\odot$ and dimensionless spin $\chi_{BH}=0.8$ typical of neutron-star merger remnants \cite{radice_binary_2018,nedora_numerical_2021,bernuzzi_mergers_2014,sekiguchi_dynamical_2016,kastaun_black_2013,foucart_black-hole-neutron-star_2012,foucart_dynamical_2017}. 
The torus has a low constant specific enthalpy of $8k_B$ per baryon, electron fraction $Y_e=0.1$, and extends from $R_{\rm in,0}=22$~km $[5M_{BH}]$ to $R_{\rm out,0}=190$~km $[43M_{BH}]$.
A weak poloidal seed magnetic field confined to the interior of the torus is added, keeping the initial magnetic-to-fluid pressure ratio below $\approx 1\times 10^{-3}$.

We employ horizon-penetrating Kerr-Schild coordinates \cite{kerr_gravitational_1963} and evolve the disk in full 3D without the use of symmetries. 
A grid hierarchy of seven nested Cartesian refinement levels centered on the black hole is used. 
The finest refinement level provides a resolution of 1.3\,km and has a diameter of 390\,km and thus contains the entire initial disk. We perform two simulations with identical setup except for the treatment of fast flavor conversions: one simulation neglecting fast conversions (`\texttt{NFC}') and one accounting for fast conversions (`\texttt{FC}').

%%%%%%%%% Flavor conversion formalism %%%%%%%%%

\textit{Flavor conversion formalism.---}Following \cite{sarikas_suppression_2012}, we describe local neutrino dynamics using the density matrix in the weak-interaction basis, split into trace and trace-less parts,
\begin{equation}
	\varrho_\nu = \frac{f_{\nu_e}+f_{\nu_X}}{2} I +\frac{f_{\nu_e}-f_{\nu_X}}{2}
	\left(\begin{matrix}
  s& S \\
  S^*& -s
\end{matrix}\right),
\end{equation}
where $f$ is the initial occupation number, $S$ is the complex $\nu_e\nu_X$ coherence, and $s$ is real with $s^2+|S|^2=1$. 
We use natural units ($c=G=\hbar=1$) and adhere to the flavor isospin convention; the density matrix for anti-neutrinos is $-\varrho_\nu$. 
In a test volume above the neutrinosphere within which neutrinos can be assumed to be free-streaming, evolution of $\rho_\nu$ is governed by the von Neumann equation $iv^\mu \partial_\mu\rho_\nu = [H,\rho_\nu]$, with $v^\mu=(1,\boldsymbol{v})$ being the four-velocity of neutrinos. 
The Hamiltonian containing vacuum oscillations is given by $H = M^2/2E + H_{\rm mat} + H_\nu$ where $M$ and $E$ denote the neutrino mass matrix and energy \cite{fukuda_evidence_1998,ahmad_measurement_2001}.
$H_{\rm mat} = -v^\mu\Lambda_\mu\frac{\sigma_3}{2}$ is the MSW neutrino-matter interaction \cite{wolfenstein_neutrino_1978,mikheyev_resonance_1985,mikheev_neutrino_1986}, where $\sigma_3$ is a Pauli matrix and $\Lambda_\mu = \sqrt{2}G_{\rm F} (n_{e^-} - n_{e^+})u_\mu$ is the matter potential, with $G_{\rm F}$, $n_{e^\pm}$, and $u_\mu$ being the Fermi constant, the electron/positron number densities, and their four velocities, respectively. 
$H_\nu = -(\sqrt{2}/(2\pi)^3)G_{\rm F}\int v^\mu v_\mu' \rho_\nu' E'^2 {\rm d} E' {\rm d}\Omega'$ is the neutrino self-interaction \cite{fuller_resonant_1987,notzold_neutrono_1988,fuller_can_1992,pantaleone_dirac_1992}, which renders the equation of motion non-linear; ${\rm d} \Omega$ is the solid angle element in direction $\boldsymbol{v}$.

Neutrinos are created as flavor eigenstates ($S=0$); to describe the onset of fast conversions in a fluid element, we can linearize the von Neumann equation assuming $|S|\ll 1$. 
Given the ultra-short time ($\sim\!\text{ns}$) and length scales ($\sim\!\text{cm}$) of fast conversions, we neglect vacuum oscillations ($\sim\!\text{km}$), which makes the equation of motion energy-independent. 
We follow a dispersion-relation ansatz as in Ref.~\cite{izaguirre_fast_2017} and insert normal modes $S_{\boldsymbol{v}}(t,\boldsymbol{r})=Q_{\boldsymbol{v}}(\tilde{\varpi},\tilde{\boldsymbol{k}})\exp[-i(\tilde{\varpi} t - \tilde{\boldsymbol{k}}\cdot\boldsymbol{r})]$ into the linearized von Neumann equation to obtain a necessary criterion for fast conversions,
\begin{equation}
	v^\mu k_\mu Q_{\boldsymbol{v}} + \int {\rm d}\Omega' \;v^\mu v'_\mu G_{\boldsymbol{v}'}Q_{\boldsymbol{v}'} = 0.
\end{equation}
Here, $k_\mu = (-\varpi, \boldsymbol{k})\equiv (-\tilde{\varpi},\tilde{\boldsymbol{k}}) - \Lambda_\mu - \Phi_\mu$ captures a constant shift due to matter interactions,
\begin{equation}
	G_{\boldsymbol{v}} = \frac{\sqrt{2}}{(2\pi)^3}G_F\int {\rm d} E\,E^2\left[f_{\nu_e}(E,\boldsymbol{v})-f_{\bar{\nu}_e}(E,\boldsymbol{v})\right],
\end{equation}
and $\Phi_\mu \equiv \int {\rm d}\Omega \, G_{\boldsymbol{v}} v_\mu$
(we assume $f_{\nu_X} = f_{\bar{\nu}_X}$ as $\nu_{\mu,\tau}$ are created via pair annihilation and plasmon decay in our simulations). 
Defining the polarization vector $a_\mu \equiv - \int {\rm d}\Omega\, v_\mu G_{\boldsymbol{v}} Q_{\boldsymbol{v}}$, the evolution equation reduces to $\Pi^{\mu\nu}a_\mu = 0$, where
\begin{equation}
	\Pi^{\mu\nu} = \eta^{\mu\nu} - \int {\rm d}\Omega\, G_{\boldsymbol{v}} \frac{v^\mu v^\nu}{\varpi-\boldsymbol{k}\cdot\boldsymbol{v}},
\end{equation}
with $\eta_{\mu\nu}={\rm diag}(-1,1,1,1)$. 
Fast conversions occur when $\det \Pi^{\mu\nu}=0$ admits complex roots. 

In our simulations, we solely focus on $\boldsymbol{k}=0$. 
Should there be additional unstable modes with $\boldsymbol{k}\ne 0$, the overall effect of fast conversions on outflow properties would only be enhanced \footnote{At submission of this paper, we were made aware of a similar suggestion in the Newtonian case \cite{dasgupta_simple_2018,abbar_searching_2020}}. 
This approach is also motivated by our observation of ubiquitous $\boldsymbol{k}=0$ unstable modes around post-merger disks, such that the effect of additional unstable modes would be minute. 
Given the extremely small time and length scales of this instability (sub-grid scales), we employ the Equivalence Principle to derive the general-relativistic version of the dispersion relation, 
\begin{equation}
	\det \left[\varpi g^{\mu\nu} - \sqrt{2}G_F \left(M^{\mu\nu}_{\nu_e}-M^{\mu\nu}_{\bar{\nu}_e} \right)\right]=0, \label{eqn:dispersion_GR}
\end{equation}
where
\begin{equation}
    M^{\mu\nu}_s\equiv \frac{1}{(2\pi)^3} \int E^2 {\rm d}E {\rm d}\Omega\;f_{s} v^\mu v^\nu
\end{equation}
is the two-moment of the neutrino distribution function and $s=\{\nu_e, \bar{\nu}_e\}$. 
In the 3+1 split of spacetime into non-intersecting spacelike hypersurfaces employed in our simulations, $M^{\mu\nu}$ may be expressed as 
\begin{equation}
	M^{\alpha\beta} = \int \frac{{\rm d}\nu}{E}
	\left( E_{(\nu)}n^\alpha n^\beta +  F_{(\nu)}^{(\alpha} n^{\beta)} + P_{(\nu)}^{\alpha\beta}\right),
\end{equation}
where $n^\mu = (\alpha^{-1}, -\alpha^{-1}\beta^i)$ is the time-like unit normal vector that characterizes the hypersurfaces and defines the Eulerian observer, $\alpha$ denotes the lapse function, and $\beta^i$ a spatial shift vector.
Furthermore, $E_{(\nu)}$, $F_{(\nu)\alpha}$, and $P_{(\nu)\alpha\beta}$ are the zero, first, and second moment of the neutrino radiation energy density fields as seen by the Eulerian observer with frequency $\nu=E/2\pi$ in the fluid rest frame \cite{shibata_truncated_2011}.

Equation~\eqref{eqn:dispersion_GR} leads to a quartic equation in $\varpi$ with real coefficients. 
In order to monitor the onset of fast conversions, we calculate the maximum growth rate $\omega\equiv \max|\mathrm{Im}(\varpi)|$ among all roots at a given grid point (these exist in complex-conjugate pairs). To be consistent with sub-grid scales, we require growth times $\omega^{-1} < \omega_{\rm crit}^{-1}\equiv 10^{-7}$\,s much smaller than our time step of $\Delta t\approx 10^{-6}$\,s. If $\omega > \omega_{\rm crit}$, we assume a flavor instability locally develops and leads to full mixing of neutrino flavors given that the instability proceeds on ultra-short timescales (approximate flavor equipartition \cite{sawyer_speed-up_2005,2009PhRvD..79j5003S,sawyer_neutrino_2016,hansen_chaotic_2014,mirizzi_self-induced_2015,dasgupta_temporal_2015,capozzi_self-induced_2016,martin_nonlinear_2019}). 
We then set $f_{\nu_e}=f_{\nu_\mu}=f_{\nu_\tau}$ in our M1 scheme, and similarly for anti-neutrinos, which ensures conservation of lepton number; this is justified by the fact that the weak matter interaction timescales $\tau_{\nu}$ obey the following separation of timescales in our simulations: $\omega^{-1}\ll \Delta t \ll \tau_{\nu}$. Equipartition is assumed here with the caveat that such maximal mixing may overestimate the impact of flavor conversions on ejecta composition; the detailed outcome of the instability still requires further investigation \cite{padilla-gay_multi-dimensional_2021,Bhattacharyya_vdependent_2021,Richers_sim_2021}. Future parametrizations of the outcome of 3D fast flavor conversions in terms of local properties of the neutrino radiation fields can be included in our formalism.

%%%%%%%%% Results %%%%%%%%%

\textit{Results.---}In a brief initial relaxation phase, the magneto-rotational instability (MRI) \cite{1991ApJ...376..214B,1998RvMP...70....1B,2003ARA&A..41..555B,2011ARA&A..49..195A} is initiated in the disk due to differential rotation. 
A quasi-stationary accretion state is established by $20$\,ms, which we regard as the actual initial data of our simulations. This steady turbulent state is indicated by a constant ratio of the electromagnetic energy to internal energy of the disk over the remainder of the simulation phase, as well as by a fully operational magnetic dynamo, which manifests itself in a `butterfly diagram'---cyclic migration of toroidal magnetic fields of alternating polarity to higher latitudes \cite{siegel_three-dimensional_2017,siegel_three-dimensional_2018,de_igniting_2020}. 
The fastest-growing mode of the MRI is resolved by more than ten grid points once in the stationary regime. 

\begin{figure}[tb]
\centering
\includegraphics[width=0.45\textwidth]{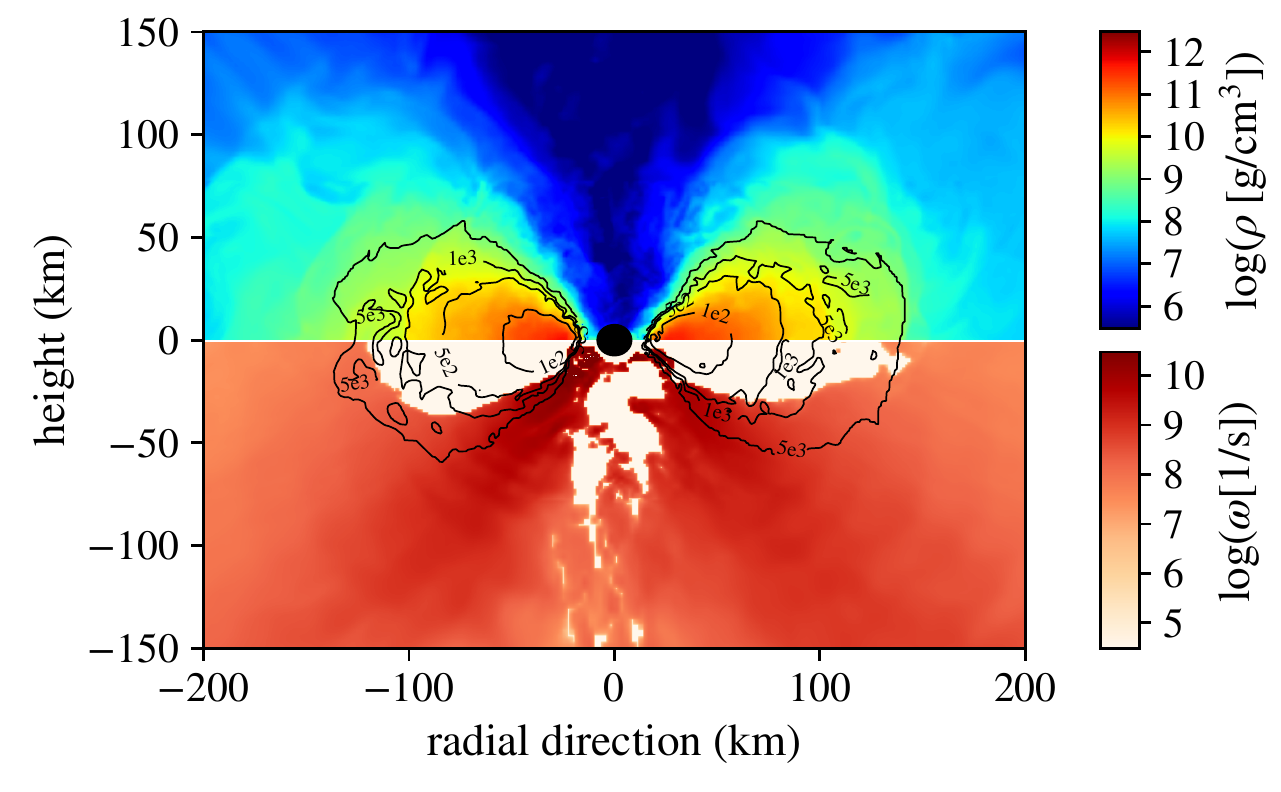}
\includegraphics[width=0.45\textwidth]{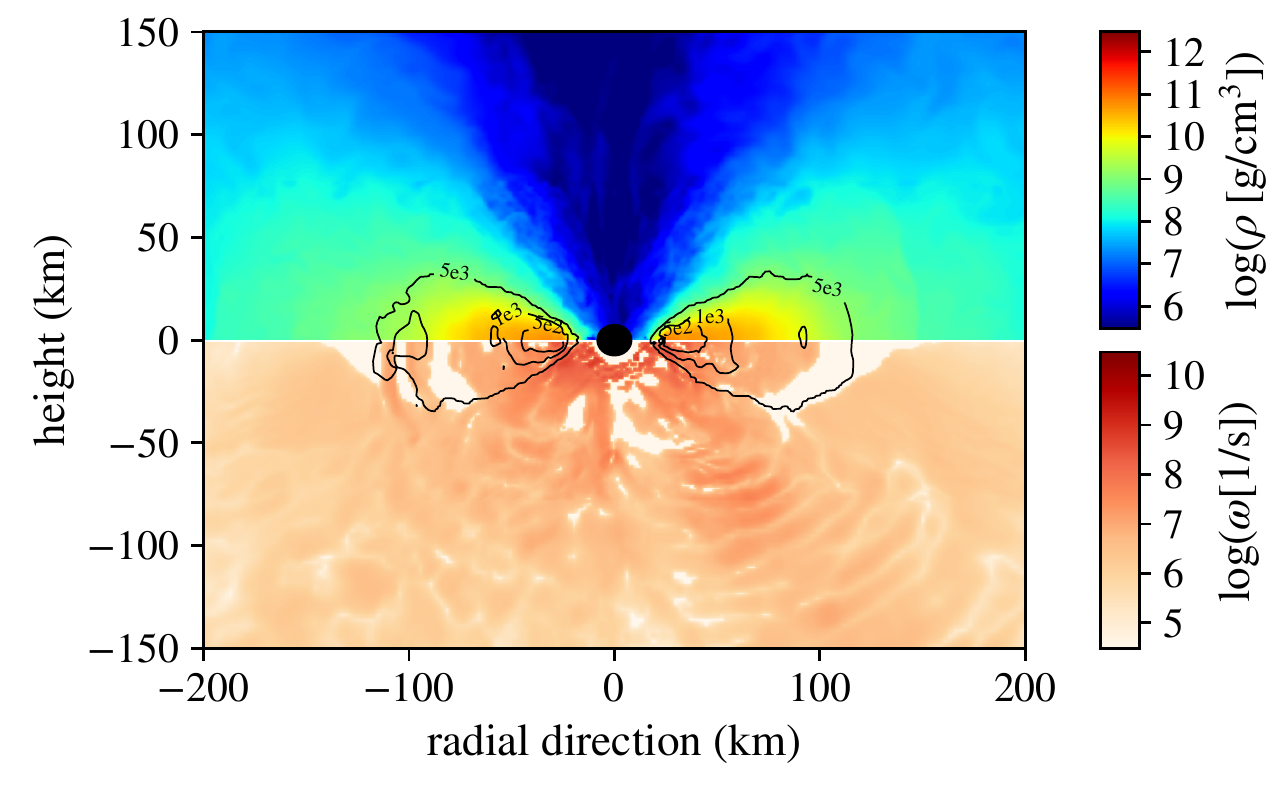}
 \caption{Snapshots of rest-mass density $\rho$ and growth rate $\omega$ for fast flavor conversions in the meridional plane at 30\,ms (top) and $300$\,ms (bottom) for the \texttt{FC} run. White regions indicate stability against fast flavor conversions. Contours delineate the mean free path $1/\kappa$ at $100$, $500$, $1000$, $5000$\,km for $\bar{\nu}_e$ in the energy range of 10--20\,MeV.
  }
  \label{fig:snapshot_rho_tau}
\end{figure}

\begin{figure}[tb]
\centering
\includegraphics[width=0.45\textwidth]{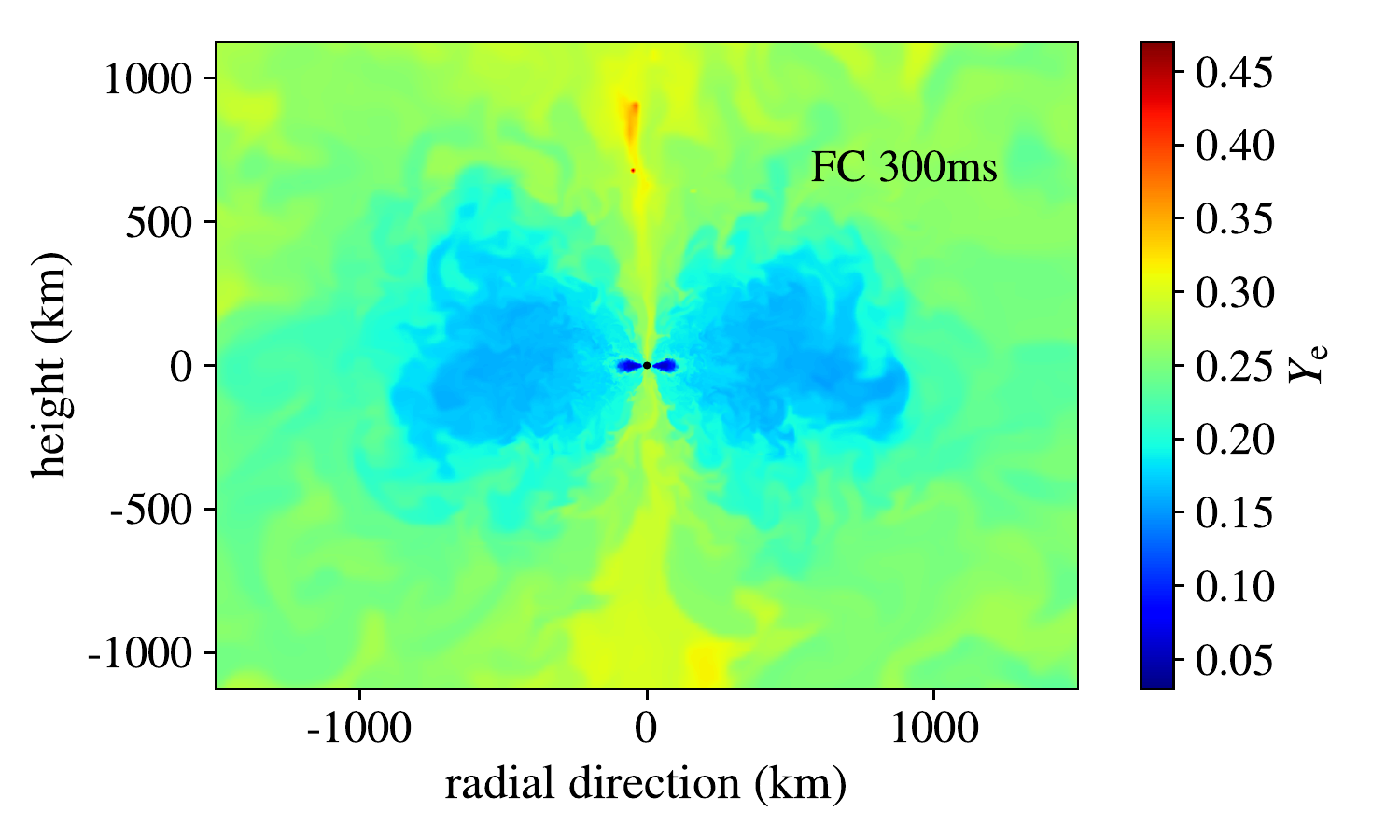}
\includegraphics[width=0.45\textwidth]{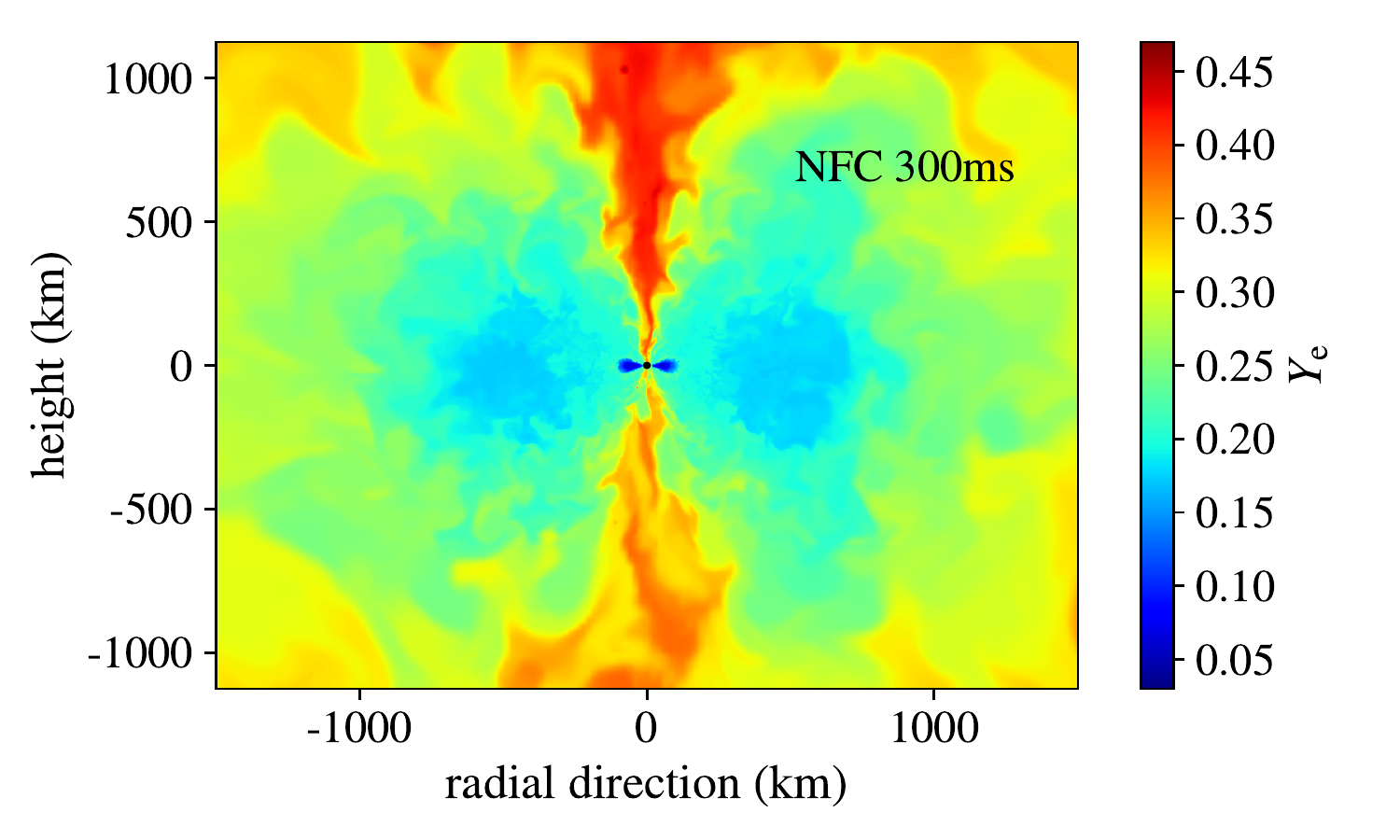}
 \caption{Snapshots of the proton fraction in the meridional plane at $300$\,ms for the \texttt{FC} run (top) and the \texttt{NFC} run (bottom), showing the emergence of an angular and radially dependent composition profile of the outflows. Fast flavor conversions lead to significantly more neutron-rich outflows. 
  }
  \label{fig:snapshot_Ye}
\end{figure}

The early stationary state of the disk is shown in the top panel of Fig.~\ref{fig:snapshot_rho_tau}. 
MRI mediated turbulence leads to angular momentum transport and sets a self-consistent accretion rate at $\dot{M}\approx 1\,M_\odot\,\text{s}^{-1}$. %after relaxation. 
As a result of angular momentum transport, the disk viscously spreads and the accretion rate decreases over time, remaining at high values of $>\!0.1\, (\text{few}\times 10^{-2}) \,M_\odot\,\text{s}^{-1}$ over the first $100$ ($400$)\,ms. 
This accretion rate being above the critical threshold $\dot{M}_{\rm ign}\sim 10^{-3} M_\odot\,\text{s}^{-1}$ for weak interactions to become energetically significant \cite{chen_neutrino-cooled_2007,metzger_time-dependent_2008,de_igniting_2020}, results in a changing composition of the accretion flow and disk outflows over time. 
While the outer parts of the disk protonize due to $\bar{\nu}_e$ emission by positron captures on neutrons ($Y_e>0.1$; Fig.~\ref{fig:snapshot_Ye}), which dominate over opposing electron captures on protons, the electrons in the inner high-density part of the disk become degenerate and a self-regulation mechanism \cite{chen_neutrino-cooled_2007,siegel_three-dimensional_2017} ensures the accretion flow stays strongly neutron-rich over viscous timescales ($Y_e\approx 0.1$). 

Powerful disk outflows are launched by imbalanced viscous heating at higher latitudes where neutrino cooling becomes energetically subdominant \cite{siegel_three-dimensional_2018}. The composition in these outflows is influenced by re-absorption of neutrinos emitted by the disk via $\bar{\nu}_e + p \rightarrow n + e^+$ and $\nu_e + n \rightarrow p + e^-$, which is accounted for by our M1 radiation transport scheme. Reabsorption and enhancement of $Y_e$ is particularly strong in the low-density polar regions (cf.~Figs.~\ref{fig:snapshot_rho_tau} and \ref{fig:snapshot_Ye}). Although M1 schemes tend to somewhat overenhance $Y_e$ compared to Monte-Carlo based approaches in polar regions \cite{Foucart_2020}, this is not a concern regarding outflow properties in the present study, since only $\approx\!0.2\%$ of the total ejecta originate from within $\approx 26^\circ$ about the polar axis. For the same reason, $\nu_e\bar{\nu_e}$ annihilation in the polar regions can be safely neglected, as done here.
In addition to angular dependence, we find a radially dependent $Y_e$ profile, which translates into a radial lanthanide gradient once the r-process proceeds \cite{lippuner_r-process_2015} (Fig.~\ref{fig:snapshot_Ye}). 
This is the result of decreasing neutrino emission from the disk and re-absorption of neutrinos in the outflows as the disk viscously spreads and its midplane density drops over time; the self-regulated inner disk injects increasingly neutron-rich material into the outflows. 
The radial lanthanide gradient is more pronounced in the absence of fast conversions (\texttt{NFC} run), as these suppress the $\nu_e$ and $\bar{\nu}_e$ fluxes.

Once the disk reaches a quasi-stationary state, fast flavor conversions emerge above the energy-dependent neutrinospheres in the disk-corona transition where neutrinos start to decouple and stream freely (Fig.~\ref{fig:snapshot_rho_tau}). 
Such conversions are ubiquitous, rendering essentially all of the disk vicinity into an unstable region with typical instability growth times up to $\omega^{-1}\sim 0.1$\,ns, in agreement with previous semi-analytic predictions \cite{wu_fast_2017,wu_imprints_2017}. 
At later times ($\sim\!300$\,ms), the instability region shrinks overall and expands into the disk as the density in the disk and outflows decreases, with slower growth rates on an increasing timescale $1-100$\,ns. 
Our results do not depend on $\omega_{\rm crit}$, as long as it is small enough to allow an extended instability region to emerge.
The composition of outflows only weakly depends on the radial size of the instability region, as the neutrino flux and the composition is mostly set in the densest part deep in the outflow, whereafter $Y_e$ starts to `freeze out'.

\begin{figure}[tb]
\centering
\includegraphics[width=0.45\textwidth]{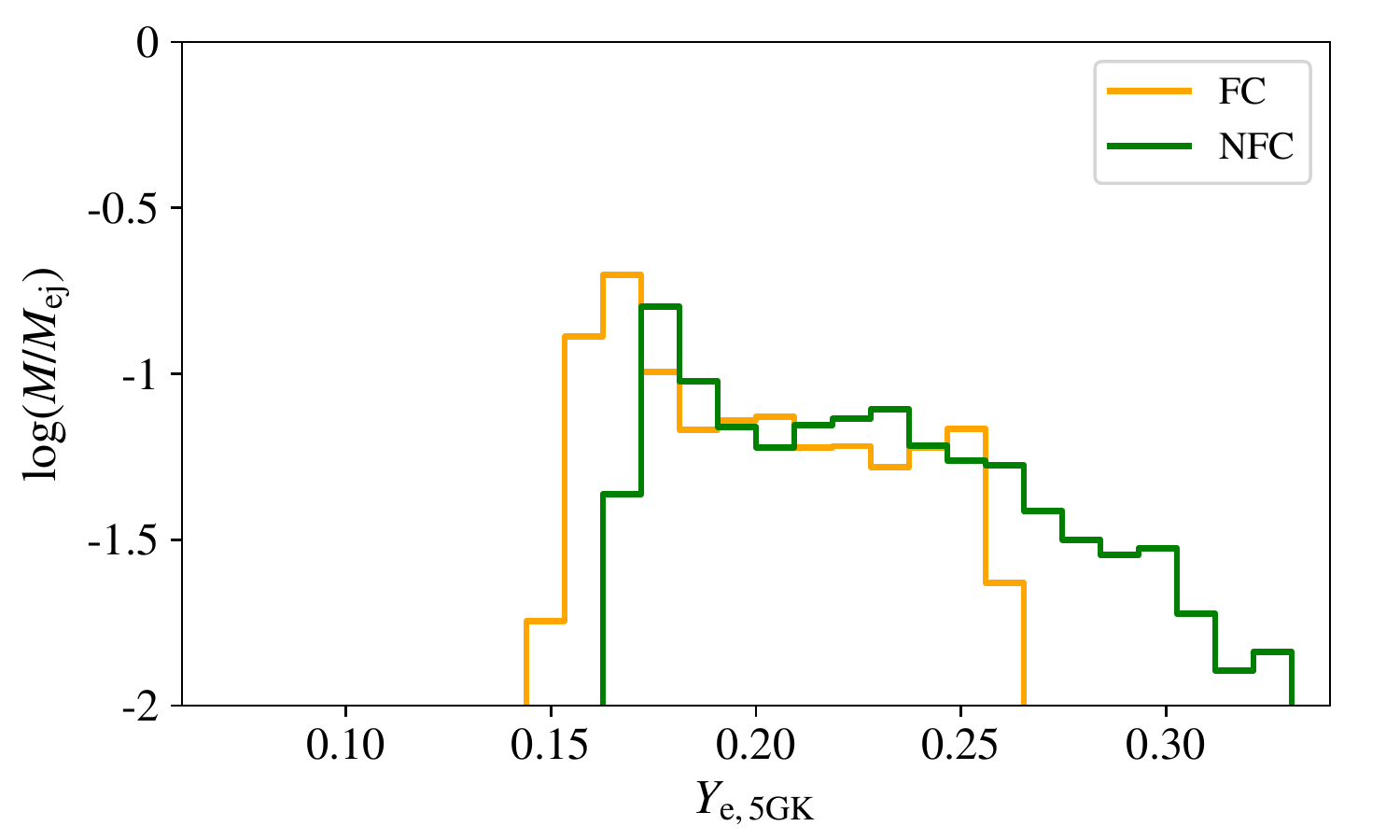}
\includegraphics[width=0.45\textwidth]{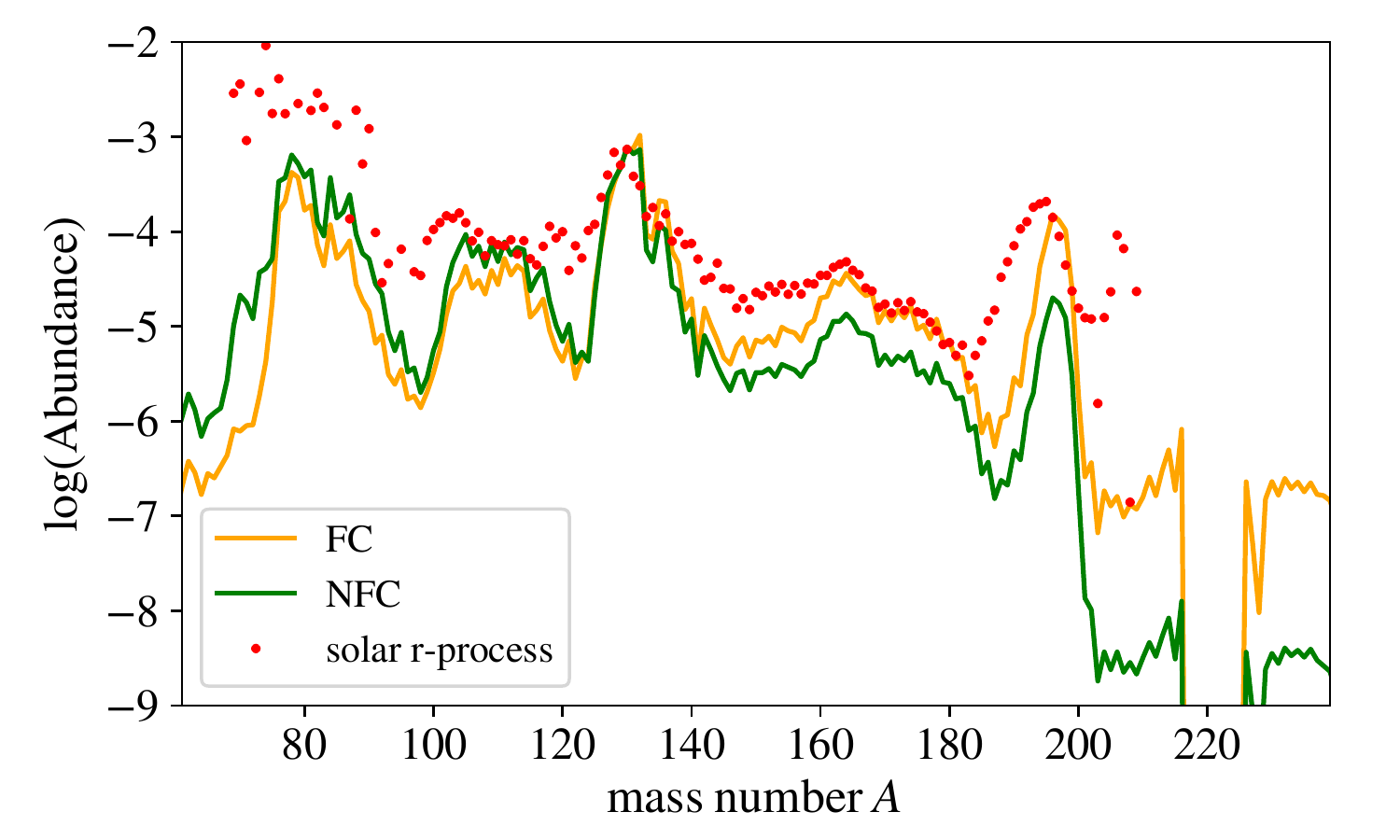}
 \caption{Top: Distribution of proton fraction for unbound tracer particles at the onset of neutron-capture reactions at 5\,GK, normalized by total ejecta mass. Fast conversions truncate the high-$Y_e$ tail. 
 Bottom: Final nucleosynthetic abundances at $10^9$\,s from reaction network calculations for the \texttt{FC} and \texttt{NFC} runs, compared to solar abundances \cite{arnould_r-process_2007}. Fast conversions boost lanthanide abundances (atomic mass number $136\lesssim A \lesssim 176$), while keeping similar abundances for lighter elements.}
  \label{fig:nucleosynthesis}
\end{figure}

Outflow properties are recorded with $10^5$ passive tracer particles initially placed in the disk proportional to conserved rest-mass density. 
Trajectories of tracers that have reached a distance $>700$\,km by the end of the simulations and that are unbound according to the Bernoulli criterion ($-hu_t>1$, where $u_t$ is the time component of the four-velocity and $h$ the specific enthalpy) are employed as input to nuclear reaction network calculations with \texttt{SkyNet} \cite{lippuner_skynet:_2017} to determine the resulting r-process abundances (ignoring tracers ejected during the initial relaxation phase).
Network calculations are performed accounting for neutrino absorption, starting in nuclear statistical equilibrium at $T = 7$\,GK. 
The network calculations consider 7843 nuclides and 140,000 nuclear reactions, using strong reaction rates from the JINA REACLIB database \cite{Cyburt_reaclib_2010}, with inverse rates derived from detailed balance, weak rates from Refs.~\cite{fuller_weakrates_nulib,oda_weakrates_nulib,langanke_weakrate_nulib} where available and from REACLIB otherwise, spontaneous and neutron-induced fission rates from Refs.~\cite{Frankel_fission_1947,Panov_rates_2010,Mamdouh_fission_2001,wahl_fission_2002}, and the REACLIB nuclear masses and partition functions (a combination of experimental data where available and data from the finite range droplet macroscopic model (FRDM) \cite{MOLLER20161}).
The $\nu_e$ and $\bar{\nu}_e$ fluxes for absorption are obtained from the simulations by fitting a Dirac-Fermi distribution to match the total local number density and average energy for each species and are extrapolated beyond the evolution time by power laws.
We note that the projected total unbound mass of $\approx\!0.026\,M_\odot$ (\texttt{FC}) and $\approx\!0.03\,M_\odot$ (\texttt{NFC}) as well as the mass-averaged velocity of $\approx\!0.1c$ of the ejecta only mildly differ between the two runs, since neutrinos play a minor role in setting the outflow energetics for these disk winds.

The imprint of fast flavor conversions on the initial conditions for r-process nucleosynthesis are shown in the top panel of Fig.~\ref{fig:nucleosynthesis}. 
The instability decreases the flux of $\nu_e$ and $\bar{\nu}_e$ due to mixing into the other less luminous flavors and thus leads to less neutrino re-absorption in the outflows, reducing the overall $Y_e$. 
This is echoed by Fig.~\ref{fig:snapshot_Ye} and consistent with previous estimates \cite{wu_imprints_2017}. 
The final r-process abundances show strongly enhanced lanthanide and beyond-2nd-peak mass fractions in the presence of fast conversions (Fig.~\ref{fig:nucleosynthesis}; Tab.~\ref{table:massfrac}), in broad agreement with solar abundances.

\begin{table}
\begin{tabular}{c|>{\centering\arraybackslash}m{0.10\textwidth}|>{\centering\arraybackslash}m{0.10\textwidth}|>{\centering\arraybackslash}m{0.10\textwidth}} 
\hline
 \texttt{SkyNet} run & $X_{\rm 2nd}$ & $X_{\rm 3rd}$ & $X_{\rm La}$ \\
 \hline
\texttt{FC} & 0.631 & 0.134 & 0.097\\
\texttt{NFC} & 0.709 & 0.023 & 0.049\\
solar r-process & 0.347 & 0.183 & 0.139\\
\hline
\end{tabular}
\caption{Mass fractions of the 2nd ($125\le A \le 135$) and 3rd ($186\le A \le 203$) r-process peak as well as of lathanides in the disk outflows simulated with and without accounting for fast conversions. Solar abundances are also listed \cite{arnould_r-process_2007}.}
\label{table:massfrac}
\end{table}

%%%%%%%%% Conclusion %%%%%%%%%

\textit{Conclusion.---}In contrast to previous simulations at $\dot{M}\lesssim \text{few}\times10^{-2}\,M_\odot\,\text{s}^{-1}$ \cite{siegel_three-dimensional_2017,de_igniting_2020,fernandez_long-term_2019}, the regime of up to $\dot{M}\approx 1\,M_\odot\,\text{s}^{-1}$ initially probed here reveals outflows with a radially dependent proton fraction. 
This translates into `bluer' kilonova emission at early times, followed by a `red' component of lanthanide-rich material ejected later. 
3D kilonova radiation transport calculations will be required to translate the resulting %angular and radially dependent 
lanthanide distribution into detailed predictions for kilonova emission.

Our results show that neutrino fast flavor conversions can boost the lanthanide and 3rd-peak r-process mass fractions in outflows from high-$\dot{M}$ disks (at the typical to upper end expected for post-merger disks) to a level comparable to solar abundances. 
This provides strong evidence for massive post-merger disks being a major production site for heavy r-process elements. 
The total estimated ejected mass scale of $\approx\!0.03\,M_\odot$ as well as the other characteristics of the outflow (composition and velocity of $\sim\!0.1c$) suggest that a similar accretion scenario may have occurred in GW170817. 
Lanthanide enhancement as found here may be required to compensate for the `protonizing presence' of a remnant neutron star of moderate lifetime, which has been argued is necessary to explain the `blue' component of the GW170817 kilonova \cite{metzger_magnetar_2018,siegel_gw170817_2019,metzger_kilonovae_2019}. 
Furthermore, the effect of fast conversions on ejecta composition demonstrated here impacts equation-of-state constraints derived from kilonova colors and will thus need to be taken into account in future analyses.

The boost in beyond-2nd-peak r-process elements due to flavor conversions will make it more challenging to obtain appreciable `blue' kilonova emission from neutron star--black hole mergers. 
It may also be of importance in explaining actinide boost stars \cite{mashonkina_hamburgeso_2014} from very neutron-rich outflows of such mergers. 
Furthermore, a similar process may be operating in collapsars to significantly increase the critical accretion rate below which lanthanides can be formed \cite{siegel_collapsars_2019,2019PhRvD.100b3008M}. 
Our formalism may also be employed in simulations of ordinary core-collapse supernovae to explore whether fast flavor conversions strengthen or weaken the explosion, although extensions to $k\ne0$ may be required to fully capture the instability \cite{nagakura_fast-pairwise_2019}.

Future studies may focus on including other flavor instability modes $\boldsymbol{k}\ne0$. This will unlikely alter our conclusions due to the $\boldsymbol{k}=0$ mode being ubiquitous and already capturing the effect on nucleosynthesis over times in which most of the mass outflow is generated. 
The assumption of maximal mixing in flavor space requires further investigation \cite{Bhattacharyya_vdependent_2021,Richers_sim_2021}; it may overestimate the effect of fast conversions on ejecta composition.
We note that other conversions such as into a sterile neutrino would give rise to similar effects.

%%%%%%%%%%%%%%%%%%%%%%%%%%%%%%%%%%%%%%%%%
%% ACKNOWLEDGEMENTS 
%%%%%%%%%%%%%%%%%%%%%%%%%%%%%%%%%%%%%%%%%

\acknowledgments

The authors thank B.~Metzger, M.-R.~Wu, J.~Miller for discussions and comments. This research was enabled in part by support provided by SciNet (www.scinethpc.ca) and Compute Canada (www.computecanada.ca). 
XL is supported by the Natural Sciences and Engineering Research Council of Canada (NSERC), funding reference \#CITA 490888-16 and the Jeffrey L. Bishop Fellowship.
DMS acknowledges the support of the Natural Sciences and Engineering Research Council of Canada (NSERC). Research at Perimeter Institute is supported in part by the Government of Canada through the Department of Innovation, Science and Economic Development Canada and by the Province of Ontario through the Ministry of Colleges and Universities.

\bibliographystyle{apsrev4-2}
\bibliography{ms}% Produces the bibliography via BibTeX.

\end{document}